\documentclass[a4paper]{article}
\pdfoutput=1
\usepackage{jheppub}
\usepackage[utf8]{inputenc}
\usepackage[T1]{fontenc}
\newcommand{\la}{\lambda_1}
\newcommand{\lb}{\lambda_2}
\newcommand{\lc}{\lambda_3}
\newcommand{\ld}{\lambda_4}
\newcommand{\lp}{\lambda_5}

\newcommand{\lczp}{\lambda_{345}}
\newcommand{\ba}{\begin{array}}
\newcommand{\ea}{\end{array}}
\newcommand{\g}{\,\mathrm{GeV}}
\newcommand{\m}{M_{H^{\pm}}}
\newcommand{\f}{\varphi}

\newcommand{\be}{\begin{equation}}
\newcommand{\ee}{\end{equation}}
\newcommand{\veff}{V_{\mathrm{eff}}^{(1)}}
\usepackage{amsthm}
\usepackage{amsfonts}
\allowdisplaybreaks[4]
\title{Inert scalars and vacuum metastability\\ around the~electroweak scale}
\author{Bogumi{\l}a \'{S}wie{\.z}ewska}
\affiliation{Faculty of Physics, University of Warsaw, Pasteura 5, 02-093 Warsaw, Poland}
\emailAdd{Bogumila.Swiezewska@fuw.edu.pl}

\abstract{
We analyse effective potential around the electroweak (EW) scale in the  Standard Model (SM)  extended with a heavy scalar doublet. We show that the additional scalars can have a~strong impact on vacuum stability.  Although the additional heavy scalars may improve the~behaviour of~running Higgs self-coupling at large field values, we prove that they can destabilise the~vacuum due to  EW-scale effects. A new EW symmetry conserving minimum of the effective potential can appear rendering the electroweak symmetry breaking (EWSB) minimum meta- or unstable. However, for the case of the inert doublet model (IDM) with a 125~GeV Higgs boson we demonstrate that the parameter space region where the vacuum is meta- or unstable cannot be reconciled with the~constraints from perturbative unitarity, electroweak precision tests (EWPT) and dark matter relic abundance measurements.}

\arxivnumber{1503.07078}

\begin{document}

\maketitle
\flushbottom
%~~~~~~~~~~~~~~~~~~~~~~~~~~~~~~~~~~~~~~~
\section{Introduction}
%~~~~~~~~~~~~~~~~~~~~~~~~~~~~~~~~~~~~~~~
In the summer of 2012 the Higgs boson was discovered~\cite{Atlas-Higgs, CMS-Higgs}, and with the measurement of its mass the issue of vacuum stability gained a lot of attention. State of the art computations show that the~SM  vacuum is metastable --- it is not a global minimum of the potential but its lifetime is extremely long~\cite{Degrassi:2012, Buttazzo:2013, Spencer-Smith:2014, Kobakhidze:2014}. However, this is not the final answer to the question of vacuum stability because the SM does not describe all phenomena that we know, and some beyond SM (BSM) theories are necessary. New BSM interactions can modify vacuum structure of the potential and change the lifetime of  the EWSB vacuum~\cite{Branchina:2013prl, Branchina:2014jhep, Branchina:2014sher, Lalak:2014, Greenwood:2008}.\footnote{In principle also gravity may affect vacuum stability, for some attempts to include gravitational effects see refs.~\cite{Loebbert:2015, Burda:2015}.} 

Well studied extensions of the SM are two-Higgs-doublet models (2HDM). The IDM~\cite{Ma:1978, Cao:2007, Barbieri:2006} is  a~special  $\mathbb{Z}_2$-symmetric 2HDM providing a viable candidate for the DM particle~\cite{LopezHonorez:2006,Dolle:2009, Honorez:2010,Sokolowska:2011,Gustafsson:2012}. Moreover, in its spectrum it has a SM-like Higgs boson which is in agreement with recent experimental data~\cite{Swiezewska:2012,Krawczyk:2013jhep, Arhrib:2013, Goudelis:2013, Chakrabarty:2015}. In the~present paper we  study how the~additional scalars affect the vacuum structure of~the~effective potential of the IDM.

To study  stability of a vacuum state one normally starts from requiring the (effective) potential to~be bounded from below (positivity conditions). The common way of achieving this at the one-loop level is to check the tree-level positivity conditions with running couplings inserted. In presence of additional scalars the running Higgs self-coupling receives additional positive contribution, which helps to stabilise the~potential. It has been shown that indeed in the IDM the potential is stable up to higher energy scales than the SM potential~\cite{Kannike:2011, Goudelis:2013}. This is, however, not enough for stability of the EWSB vacuum since the~additional scalars can modify the structure of the effective potential introducing new minima, potentially deeper than the EWSB minimum.  The aim of the present article is to examine the structure of the potential and stability of the vacuum state around the EW scale in the presence of inert scalars. We will show that the potential can be significantly modified, and the EWSB minimum can be rendered meta- or unstable.

The paper is organised as follows. In section~\ref{sec:IDM} the model is briefly introduced. Section~\ref{sec:eff-pot} explains our use of the effective potential, and in section~\ref{sec:tunnelling} the computation of the lifetime of the~vacuum is described. The results of the paper are presented in section~\ref{sec:results}. Section~\ref{sec:conclusions} summarises the~conclusions.

%~~~~~~~~~~~~~~~~~~~~~~~~~~~~~~~~~~~~~~~
\section{IDM at tree level\label{sec:IDM}}
%~~~~~~~~~~~~~~~~~~~~~~~~~~~~~~~~~~~~~~~

The IDM is a special version of 2HDM~\cite{Ma:1978, Cao:2007, Barbieri:2006}. The most attractive feature of the IDM is that it provides a viable DM candidate which can account for the observed relic density of DM in agreement with direct detection constraints~\cite{LopezHonorez:2006,Dolle:2009, Honorez:2010,Sokolowska:2011,Gustafsson:2012, Arhrib:2013,Abe:2014}. Moreover, within the model thermal evolution of the Universe~\cite{Ginzburg:2010, Sokolowska:2011, Sokolowska:2011t} and strong electroweak phase transition~\cite{Kanemura:2004, Gil:2012, Cline:2013,  Chowdhury:2011} can be studied. With a~slight extension of the model, neutrino masses can be accounted for~\cite{Ma:2006neutrino, Gustafsson:2012neutrino,Chakrabarty:2015, Merle:2015}. Moreover, the~IDM can be constrained with the use of accelerator data, such as invisible Higgs decay branching ratios and the diphoton Higgs decay rate~\cite{Swiezewska:2012,Krawczyk:2013jhep,Arhrib:2013,Gustafsson:2012, Goudelis:2013, Chakrabarty:2015}.

Scalar interactions of two $SU(2)$ doublets in the IDM  are given by the following potential
\begin{align}
V=&-\frac{1}{2}\left[m_{11}^{2}(\phi_{S}^{\dagger}\phi_{S})+m_{22}^{2}(\phi_{D}^{\dagger}\phi_{D})\right]+\frac{1}{2}\left[\lambda_{1}(\phi_{S}^{\dagger}\phi_{S})^{2}+\lambda_{2}(\phi_{D}^{\dagger}\phi_{D})^{2}\right]\nonumber\\
&+\lambda_{3}(\phi_{S}^{\dagger}\phi_{S})(\phi_{D}^{\dagger}\phi_{D})
+\lambda_{4}(\phi_{S}^{\dagger}\phi_{D})(\phi_{D}^{\dagger}\phi_{S})+\frac{1}{2}\lambda_{5}\left[(\phi_{S}^{\dagger}\phi_{D})^{2}+(\phi_{D}^{\dagger}\phi_{S})^{2}\right].\label{potential-IDM}
\end{align}
The potential is symmetric under two $\mathbb{Z}_2$ transformations, $D: \phi_D \to -\phi_D, \ \phi_S\to \phi_S$ and $S: \phi_S\to -\phi_S, \ \phi_D\to \phi_D$,  the SM fields are assumed not to change under these transformations. We choose $D$ as a symmetry of our model. To~preserve it Yukawa interactions are set to type I (i.e. only the~$\phi_S$ doublet couples to fermions), and at~tree level a~$D$-symmetric vacuum state is considered 
\be
\langle\phi_S\rangle=\frac{1}{\sqrt{2}}\left(\begin{array}{c}0\\ v\end{array}\right), \quad\langle\phi_D\rangle=\left(\begin{array}{c}0\\ 0\end{array}\right).\label{clas-fields}
\ee
This way the whole model is $D$-invariant, and $D$ parity is a conserved quantum number.

The $\phi_S$ and $\phi_D$ doublets can be decomposed around the vacuum state in the following way
$$
\phi_S=\frac{1}{\sqrt{2}}\left(\begin{array}{c}\sqrt{2}G^+\\ v+h+iG\end{array}\right), \quad \phi_D=\frac{1}{\sqrt{2}}\left(\begin{array}{c}\sqrt{2}H^+\\ H+iA\end{array}\right),
$$
where all the fields are mass eigenstates, $G$ and $G^{\pm}$ are pseudo-Goldstone bosons, and $h$ is the~Higgs boson. The tree-level masses of the physical particles read
\begin{align}
M_{h}^{2}&=-\frac{1}{2}m_{11}^2 +\frac{3}{2}\la v^{2}=\la v^{2}=m_{11}^2,\nonumber\\
M_{H^{\pm}}^{2}&=\frac{1}{2}(-m_{22}^{2}+\lc v^{2}),\nonumber\\
M_{A}^{2}&=\frac{1}{2}(-m_{22}^{2}+\lczp^{-}v^{2}),\label{tree-masses}\\
M_{H}^{2}&=\frac{1}{2}(-m_{22}^{2}+\lczp v^{2}),\nonumber
\end{align}
where $v^2=m_{11}^2/\la$, $\lczp=\lc+\ld+\lp$, and $\lczp^-=\lc+\ld-\lp$. 

Here we use the value of $v=250.6\g$. A common way to compute $v$ is to use its relation to the Fermi constant $v^2=\frac{1}{\sqrt{2}G_F}$, which gives the value $v=246.2\g$. However, in our computations we needed exact cancelation between terms coming from the Coleman-Weinberg (CW) potential, and from the on-shell (OS) renormalisation procedure (see section~\ref{sec:eff-pot}). In the latter, the tree-level masses of $W$ and $Z$ appear, and thus taking their measured values from Particle Data Group~\cite{PDG:2014} ($M_W=80.385\g$, $M_Z=91.1876\g$), and the fine structure constant $\alpha=1/137$ as input we have to compute $v$ using the tree-level relation with these quantities, namely
$$
v=\frac{2M_W}{\sqrt{4\pi\alpha}}\sqrt{1-\left(\frac{M_W}{M_Z}\right)^2}\approx 250.6\g.
$$ 

The $h$ particle is a SM-like Higgs boson so we fix its mass to $125\g$~\cite{Aad:2015}. It has all tree-level couplings to fundamental fermions and gauge bosons equal to the respective couplings in the SM. The $D$-odd particles $A, H, H^{\pm}$ are jointly referred to as dark or inert scalars, as they do not couple to fermions at tree level. In contrast, they do interact with gauge bosons through the covariant derivative. They always appear in pairs in interaction vertices due to conservation of $D$ parity. 

The lightest neutral $D$-odd particle, $H$ or $A$, is stable and thus can play a role of the DM particle.  The two options are exactly equivalent, they differ just by the sign of the $\lp$ parameter. Here we choose $H$ as the DM candidate, and thus partially fix the mass hierarchy: $M_H<M_A, \m$. This implies that $\lp<0$, and $\ld+\lp<0$. In the light of current experimental constraints, there are two ranges of masses of DM with correct relic density: $M_H\lesssim M_W$ and $M_H\gtrsim 500\g$~\cite{LopezHonorez:2006, Honorez:2010,Gustafsson:2012, Arhrib:2013, Abe:2014}.

To parametrize IDM one can use the parameters appearing in the Lagrangian, i.e. $\la,\ldots,\lp, m_{22}^2$ ($m_{11}^2$ is fixed by eq.~(\ref{tree-masses})). Alternatively, physical parameters can be used, e.g.  $\lb$, $\lczp$, $M_H$, $M_A$, $\m$. In the following analysis we will employ the latter. The two sets of parameters are related as follows
\begin{align}
\la&=\frac{M_h^2}{v^2},\nonumber\\
\lc&=  \frac{2}{v^2} \left(\m^2-M_H^2\right) +\lczp,\nonumber\\
\ld&=\frac{1}{v^2} \left(M_H^2 +M_A^2- 2\m^2\right),\nonumber\\
\lp& = \frac{1}{v^2} \left( M_H^2 -M_A^2\right),\nonumber\\
m_{22}^2 &= -2M_H^2 + \lczp v^2.\nonumber
\end{align}

The $\lb$ parameter in general is very hard to constrain since it is the quartic coupling between the~dark scalars.\footnote{Some tree-level constraints come from the stability of the inert vacuum, see~\cite{Sokolowska:2011}.} On the other hand, $\lczp$ is proportional to the coupling between DM particles and the~$h$ boson so it significantly influences relic density of the DM, DM-nucleon scattering cross-section, and also invisible decays of the Higgs boson to the DM particles.

%~~~~~~~~~~~~~~~~~~~~~~~~~~~~~~~~~~~~~~~
\section{Effective potential\label{sec:eff-pot}}
%~~~~~~~~~~~~~~~~~~~~~~~~~~~~~~~~~~~~~~~

A vacuum state is a ground state of a theory, i.e. a state of the lowest energy. A stable vacuum state should correspond to a global minimum of the potential. A state which is a local minimum can decay (tunnel) to the global minimum, and thus is not absolutely stable. However, if it has sufficiently long lifetime (longer than the age of the Universe)  it can  also play a role of the ground state. Such configurations will be referred to as metastable vacua. An unstable minimum (with lifetime shorter than the age of the Universe) cannot constitute a present vacuum state because it would have already decayed --- such configurations will be called unstable vacua.

To examine the vacuum structure of a model we need to analyse the effective potential~\cite{Coleman:1973}. Study of vacuum stability in models with more scalar fields is a complex task as the effective potential becomes a function of multiple variables, and new minima can appear  along various directions (see e.g. analysis in refs.~\cite{Sher:1988, Gil:2012}).\footnote{In the IDM even at tree level minima with different vacuum expectation values can coexist~\cite{Ginzburg:2010}. Similarly, in the~general 2HDM simultaneous tree-level minima can occur~\cite{Barroso:2012, Barroso:2013}. Here we focus on the loop effects.} To avoid this problem but still study the impact of the~presence of additional scalars on vacuum stability, we employ a~simplified approach. Our  assumption is that the dark scalars cannot be observed in the final/initial states, i.e. they are integrated out.  Because of this approach, we focus on the heavy DM regime, where $M_H\gtrsim 500 \g$.  In this way, in the effective potential computation we only consider one classical field on external legs of the~diagrams, and  the effective potential is a~function of only one variable. Nonetheless, loop corrections from the inert scalars are included in the one-loop renormalisation process, and their contributions to the Coleman-Weinberg (CW)~\cite{Coleman:1973} potential are taken into account.  We will show that the impact of the~new heavy scalars on vacuum structure can be significant.

The one-loop effective potential is given by
\begin{align}
V_{\mathrm{eff}}^{(1)}&=V^{(0)}_{\mathrm{eff}} +\delta V_{\mathrm{CW}}+\delta V+const.\end{align}
$V^{(0)}_{\mathrm{eff}}$ denotes the tree-level effective potential
\be
V_{\mathrm{eff}}^{(0)}=-\frac{1}{4}m_{11}^{2}\f^2+ \frac{1}{8}\lambda_{1}\f^{4}, \label{tree-level-pot}
\ee
where $\f$ is a real classical field. $\delta V_{\mathrm{CW}}$ stands for the CW potential, and $\delta V$ is the counterterm potential. A constant that shifts the potential to get $\lim_{\f\to0} V(\f) = 0$ is explicitly singled out.

The CW contribution coming from the Higgs boson, Goldstone bosons, fermions (we include top and bottom quarks as the heaviest ones), gauge bosons, and inert scalars, computed in dimensional regularisation ($D=4-\epsilon$) reads
\be
\delta V_{\mathrm{CW}}=\sum_{i}\frac{f_i}{64\pi^2} M_i(\f)^4 \left[ -\frac{2}{\epsilon} +
\gamma_E - C_i+ \log\left( \frac{M_i(\f)^2}{4\pi \mu^2}\right) \right],\label{V-CW}
\ee
where $i$ runs over particle species, and $f_i$ depends on the spin, electric and colour charge of a particle ($f_h=f_H=f_G=f_A=1$, $f_{G^{\pm}}=f_{H^{\pm}}=2$, $f_t=f_b=-12$, $f_{W^{\pm}}=6$, $f_Z=3$), and $C_i=\frac{3}{2}$ for all of~the particles, except the gauge bosons, for which $C_{W^{\pm}}=C_{Z}=\frac{5}{6}$.  For the physical particles the~field dependent masses $M_i(\f)$ are obtained by substituting $\f$ instead of $v$ in the tree-level formulas for masses. The tree-level masses of the scalars are given in eq.~(\ref{tree-masses}), for gauge bosons and fermions they read
$$
M_W=\frac{gv}{2}, \quad M_Z=\frac{\sqrt{g^2+g’^2}}{2}v, \quad M_f=\frac{y_f v}{\sqrt{2}}.
$$
The field dependent masses of the Goldstones are as follows
$$
M^2_G=M^2_{G^{\pm}}=-\frac{1}{2}m_{11}^2+\frac{1}{2}\la \f^2,
$$
which of course vanish for $\f=v$. For $\f<v$ the field-dependent masses of the Goldstone bosons become negative, and the effective potential acquires an imaginary part. Recently it has been shown that the problematic Goldstone contributions can be consistently resummed, and this way the imaginary part can be removed~\cite{Martin:2014, Elias-Miro:2014}. Furthermore, it has been demonstrated that this resummation procedure has little numerical impact on the results, thus we simply ignore the imaginary contributions from the Goldstones. 

The counterterm potential $\delta V$ is obtained after the shift in the parameters of the potential $m_{11}^2 \to m_{11}^2 + \delta m_{11}^2$, $\lambda_1 \to \lambda_1 + \delta \lambda_1$, $\f^2 \to (1+\delta Z) \f^2$ is performed,
$$
\delta V= -\frac{1}{4}\left(m_{11}^2 \delta Z + \delta m_{11}^2\right) \f^2 + \frac{1}{4}\left(\lambda_1 \delta Z + \frac{1}{2} \delta \lambda_1\right) \f^4.
$$
The counterterms are defined in the on-shell renormalisation scheme. We require that the one-loop tadpole of $h$ is cancelled --- this way the tree-level value of $v$ is preserved at the one-loop level, and that the Higgs propagator has a pole at $M_h$ with a residue equal to $i$ (we follow ref.~\cite{Gil:2012}). This gives $\delta V$ in terms of the Higgs self-energy (evaluated at $M_h^2$), $\Sigma(M_h^2)$, its derivative with respect to momentum, $\Sigma'(M_h^2)$, and the tadpole, $\mathcal{T}$,
$$
\delta V= \frac{1}{4}\left[\Sigma(M_h^2) -M_h^2 \Sigma'(M_h^2)-\frac{3\mathcal{T}}{v} \right] \f^2 - \frac{1}{8v^2}\left[\Sigma(M_h^2) -M_h^2\Sigma'(M_h^2) -\frac{\mathcal{T}}{v}\right] \f^4.
$$
The expressions for $\Sigma$ and $\mathcal{T}$ are given in the appendix~\ref{app}. In the counterterms there is another source of imaginary part of the effective potential --- the loops containing the $b$ quark. This complexity signals instability of the Higgs boson, and we can simply take into account only the real part of the potential~\cite{Weinberg:1987}.

The infinities present in $\delta V$ exactly cancel the $\frac{2}{\epsilon}$ terms in $\delta V_{\mathrm{CW}}$, together with $\gamma_E - \log(4\pi\mu^2)$.  Thus the final potential is finite and $\mu$-independent.

%~~~~~~~~~~~~~~~~~~~~~~
\section{Lifetime of the vacuum\label{sec:tunnelling}}
%~~~~~~~~~~~~~~~~~~~~~~
As we will show in the following, in the IDM with heavy inert scalars EWSB minimum is not necessarily the global one. To assess whether such a state can play a role of a metastable vacuum state we have to compute its lifetime, and check whether it is longer than the age of the Universe.  %If it is not, such state would have decayed long ago, and could not be a present vacuum state. 
In the computation of the vacuum lifetime we follow the seminal papers~\cite{Coleman:1977, Callan:1977}, and the more recent ones~\cite{Branchina:2013prl, Branchina:2014jhep, Branchina:2014sher, Lalak:2014}.

To determine the lifetime of vacuum we have to find a classical trajectory, the so-called bounce solution, $\f_B$, which satisfies the following equation (in the $O(4)$-symmetric case it depends only on one variable $s=\sqrt{\vec{x}^2+x_4^2}$):
\be
\ddot{\f} + \frac{3}{s} \dot{\f}= \frac{\partial \veff(\f)}{\partial \f},\label{eq:bounce}
\ee
where dot denotes derivative with respect to $s$. The boundary conditions are: $\dot{\f}_B(0)=0$, and $\f_B(\infty)=v$. Having this solution, an approximate relative lifetime of the vacuum $\tau$  is given by (in the units of the age of the Universe $T_U$)
\be
\tau= \frac{e^{S_E}}{\f_0^4 T_U^4}.
\ee
The formula above is an approximation since quantum fluctuations around the bounce solution in the~exponential prefactor have been replaced by another dimensionful quantity, $\f_0=\f_B(0)$, see~refs.~\cite{Branchina:2014sher, Lalak:2014}. This approximation has been shown~\cite{Branchina:2014sher} to give a good estimation of the tunnelling time. The quantity $S_E$, the Euclidean action on the bounce solution $\f_B$, is given by
\be
S_E=2\pi^2 \int ds s^3 \left[\frac{1}{2}\dot{\f}_B^2(s)+\veff(\f_B(s))\right].\label{se}
\ee

The effective potential is a rather complicated function of the classical field so it is not possible to solve eq.~(\ref{eq:bounce}) analytically. Therefore, we solve it using the undershoot-overshoot method. 

\begin{figure}[ht]
\center
\includegraphics[width=.6\textwidth]{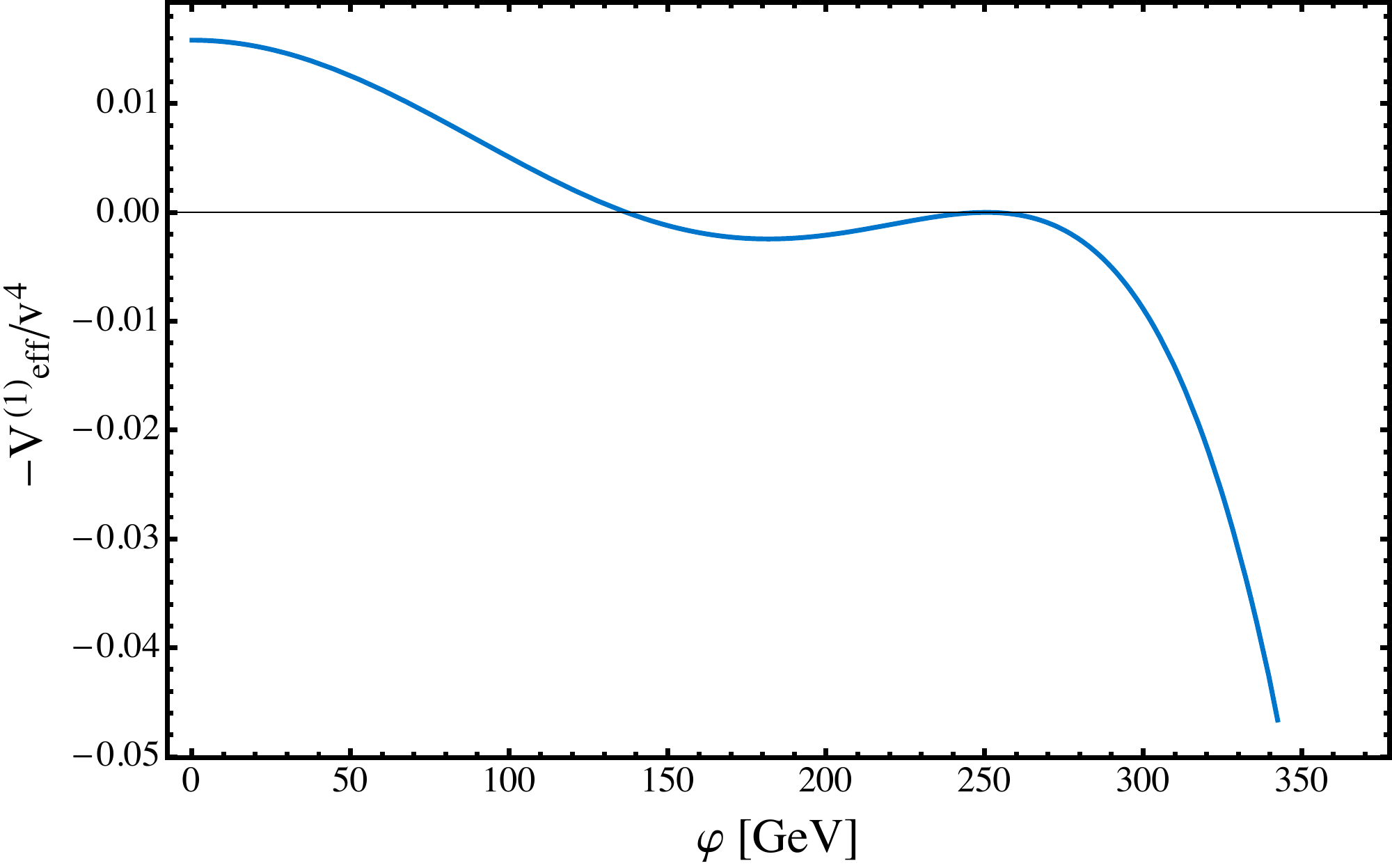}
\caption{Minus effective potential as a function of the classical field $\f$. The bounce solution corresponds to a classical trajectory of a body sliding (in presence of a friction force) from the slope on the left with zero initial velocity and stopping on the top of the lower hill on the right at infinite time $s$.\label{minus-pot}}
\end{figure}

Eq.~(\ref{eq:bounce}) can be viewed as an equation describing movement of a body in the potential $-\veff$, in the~presence of a friction force (second term of eq.~(\ref{eq:bounce})), and time denoted by $s$; see figure~\ref{minus-pot} for an exemplary shape of $-\veff(\f)$. A bounce solution corresponds to a classical trajectory of a  body  sliding down from the~slope of the higher hill (corresponding to the deeper minimum of $\veff$) with initial velocity $\dot{\f}$ equal zero, and stopping at the lower hill at infinite time $s$.\footnote{For the computation of the tunnelling time we shift the potential such that it is equal zero at $\f=v$, not at $\f=0$. Thanks to that the integrand in eq.~(\ref{se}) converges to zero for $s \to \infty$. If the vacuum energy is identified with a source of the cosmological constant, indeed it has to be very small to reproduce the observations.} The task is to find appropriate starting point: if we start to close to the peak of the bigger hill we will overshoot and the body will not stop on the other hill. If we start too far, it will not reach the top. Somewhere in between lies the correct starting point. Knowing that, we look for it using the bisection method, and solve eq.~(\ref{eq:bounce}) numerically.

%~~~~~~~~~~~~~~~~~~~~~
\section{Results\label{sec:results}}
%~~~~~~~~~~~~~~~~~~~~~
%xxxxxxxxxxx
\subsection{Effective potential and lifetime of the vacuum}
%xxxxxxxxxxx

\begin{figure}[htb]
\center
\includegraphics[width=.6\textwidth]{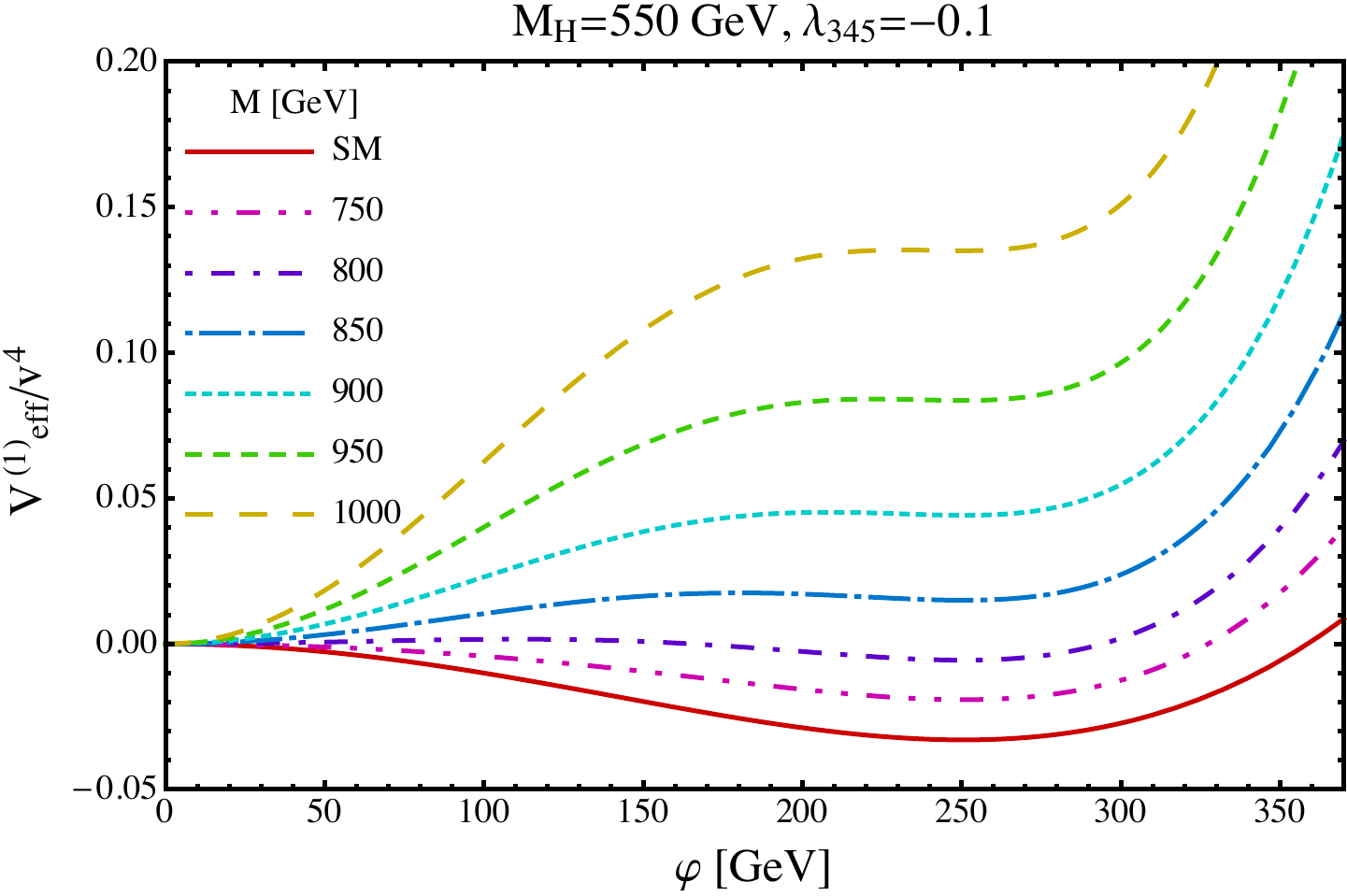}
\caption{The 1-loop OS effective potential for the IDM with heavy inert scalars integrated out. In this plot $M_H=550\g, \lczp=-0.1$, and $A$ and $H^{\pm}$ are assumed to be degenerate, with mass $M$.\label{fig:potential}}
\end{figure}

To evaluate the impact of the heavy inert scalars on vacuum stability we analyse the~structure of the~effective potential of the IDM around the EW scale. For this general discussion we fix the mass of the DM candidate to 550 GeV and $\lczp=-0.1$, as suggested by DM data (see e.g. ref.~\cite{Krawczyk:2013jhep}). The $A$ and $H^{\pm}$ particles are assumed to be degenerate, with common mass $M$. In~figure~\ref{fig:potential} the~OS effective potential for the IDM with different values of $M$ is shown. The solid line represents the~SM case (similar results were presented in ref.~\cite{Gil:2012}).

Figure~\ref{fig:potential} shows that for  lighter inert scalars the effective potential of the IDM is very close to the~SM one. While the common mass $M$ of $A$ and $H^{\pm}$ is increased (while $M_H$ is fixed), the~maximum at $\f=0$ turns to a minimum, and a maximum for $0<\f<v$ appears. Then, the~minimum at $\f=v$ becomes a~local minimum of the potential, and thus to constitute a metastable vacuum state for our model it must have long enough lifetime.

It might be surprising that the heavier the $A$ and $H^{\pm}$ scalars are, the bigger the deviation from the SM scenario is. This is because the mass of $H$ is fixed here, and increasing the splitting between $M$ and $M_H$, we increase the couplings and enter a non-decoupling regime. For $M$ and $M_H$ being close (and heavy) we are in the decoupling regime, and no significant deviation from the SM is observed.

To check whether the local minima can constitute metastable vacuum states we computed their lifetimes. We underline that we are interested here in lifetimes with respect to the tunnelling to~the EW symmetric minimum, we do not consider tunnelling to a possible minimum at very high field values. In~the~cases with $M=750,\ 800\g$ the EWSB minima are stable, their energy is lower than the energy of the EW symmetric minimum. For $M=850\g$ the tunnelling can occur but the lifetime of the EWSB vacuum is very long, $\log_{10} \tau\approx 434$ (where $\tau$ is the lifetime of the vacuum with respect to the age of the Universe). For the cases with $M=900,\ 950,\ 1000\g$ EWSB minima are highly unstable, their lifetimes are $\log_{10} \tau \approx -129, -164, -171$, respectively. Thus they cannot be considered as ground states for the~IDM. 

This shows that additional scalars can have a striking impact on the  stability of vacuum.  Although the additional heavy scalars may improve the behaviour of running Higgs self-coupling at large field values~\cite{Kannike:2011, Goudelis:2013}, they can destabilise the vacuum due to  EW-scale effects. We demonstrated this effect for the~IDM with heavy dark scalars, but one can expect similar behaviour in other models with extra scalar fields.

As was mentioned above, the interesting case of unstable EWSB minimum corresponds to relatively large splitting between $M$ and $M_H$. This suggests that “large” values of the $\lambda_i$ parameters are required. How large? For the presented cases we checked the perturbative unitarity conditions,  which constrain the parameters $\lambda_i$. In the scenarios with $M$ up to $900\g$ the conditions are fulfilled, and starting from $M=950\g$ they are violated. So parameters $\lambda_i$ required for the~meta- or unstable scenarios are rather big but still within the allowed region. In the section~\ref{sec:constraints} we confront the bounds coming from requirement of stability with other theoretical and experimental constraints in more detail to check whether meta- or instability scenarios can occur within viable parameter space of the IDM. But before that, in section~\ref{sec:pert} we study validity of the perturbative expansion of the effective potential.

%xxxxxxxxxxxxx
\subsection{Validity of the perturbative expansion\label{sec:pert}}
%xxxxxxxxxxxxx

One may ask whether the one-loop approximation of the effective potential used in this work is valid. In the OS scheme the terms of the form $\log\mu$, where $\mu$ is the renormalisation scale, cancel out between the counterterm potential and the CW contribution. As a consequence, the logarithmic terms are of the form $\log\frac{M^2(\f)}{M^2}$, where $M^2$ is the physical mass of a particle, and $M^2(\f)$ is its field dependent mass. Therefore there is no freedom of adjusting $\mu$ to make the logarithms small. 

\begin{figure}[ht]
\center
\includegraphics[width=.6\textwidth]{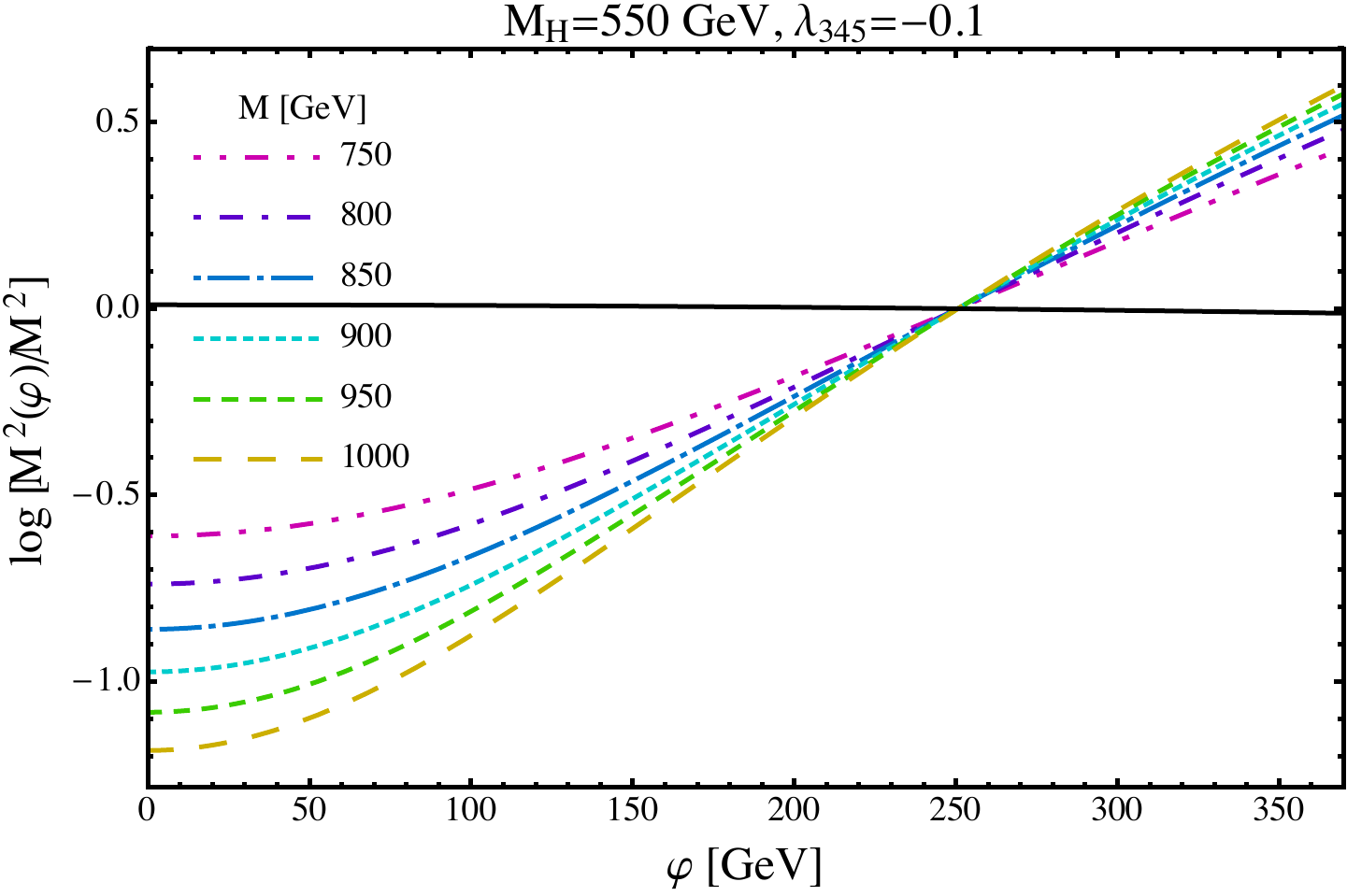}
\caption{$\log\frac{M^2(\f)}{M^2}$ as a function of $\f$. Different styles of the curves correspond to different values of $M$. The horizontal black line corresponds to $\log\frac{M_H^2(\f)}{M_H^2}$.\label{fig:logs}}
\end{figure}

The behaviour of the logarithms $\log\frac{M^2(\f)}{M^2}$ for the cases analysed above ($M_H=550\g$, $\lczp=-0.1$, $M_A=\m=M$) is shown in figure~\ref{fig:logs}. Different styles of the curves correspond to different values of $M$ (the colour coding is the same as in figure~\ref{fig:potential}). The horizontal black line corresponds to $\log\frac{M_H^2(\f)}{M_H^2}$.

It can be seen from the plot that $\log\frac{M_H^2(\f)}{M_H^2}$ is small for the whole range of $\f$. The  absolute value of the other logarithm, $\log\frac{M^2(\f)}{M^2}$, for $M\leqslant 900\g$ is less then 1 which is required for the perturbative expansion of the effective potential to be valid. For the cases with $M=950, 1000\g$ the logarithm becomes larger around $\f=0$. This could suggest breakdown of perturbative expansion, however these two cases are already excluded by perturbative unitarity, as was shown above.

One should note, that the most important point, from the perspective of this analysis, is the point $\f=v$. And at this point all the logarithms vanish, and are small around. This means that the perturbative expansion of the effective potential should be trustworthy around the electroweak minimum. Since the CW contribution vanishes around $\f=v$, the shift in the value of the potential at this point, that can be seen in figure~\ref{fig:potential}, is due to the counterterms, and the shift fixing $V^{(1)}_{\textrm{eff}}(0)=0$.

Another thing that should be taken into account is that the expansion of the effective potential is not in terms of the logarithms only, but rather in some coupling $\alpha$ times the logarithm. So the quantity $\frac{\alpha}{4\pi}\log\frac{M^2(\f)}{M^2}$ should be small~(see e.g.~\cite{Sher:1988, Camargo-Molina:2013}). It is however not so straightforward in the case of scalars to define $\alpha$, since the scalar contributions to the CW potential are not linear in $\f^4$ (in contrast to the fermionic or gauge-boson contributions). Therefore we consider separately perturbativity of the couplings in section~\ref{sec:constraints} (in terms of perturbative unitarity). Admittedly,  the couplings get rather large (within the allowed region) in the interesting cases, but as explained above, it is hard to draw final conclusions from that fact.

The standard way of improving the validity of the effective potential is using the RGEs to resum the large logarithms. However, here the source of rather big logarithms is the splitting of the scales related to masses of different particles, and therefore RGEs should not improve the situation. Thus, only a two-loop calculation could definitely show whether the one-loop potential can be trusted in the range where the logarithms become large. However, the two-loop computation  is beyond the scope of this paper.

%xxxxxxxxxxxxx
\subsection{Parameter space constraints\label{sec:constraints}}
%xxxxxxxxxxxxx

Among the relevant constraints for the IDM are
\begin{description}
\item[Perturbative unitarity.] We will assume that the eigenvalues of the scattering matrix $\Lambda_i$ fulfil $|\Lambda_i|\leqslant 8\pi$ (see e.g. ref.~\cite{Swiezewska:2012yuk}).  The allowed region in the parameter space depends on the~value of $\lb$ which is otherwise not present in our computations. The bigger the value of $\lb$, the~larger the excluded part of parameter space. Therefore in this analysis we fix $\lb$ to a~small value, $\lb=0.01$.
\item[Electroweak precision tests (EWPT).] We use the  $S$ and $T$ values from the Gfitter group, ref.~\cite{Gfitter:2014}, with $U$ fixed to 0 (the reference value of $M_h$ is 125 GeV),
\begin{align}
T&=0.10\pm0.07,\nonumber\\*
S&=0.06\pm0.09,\nonumber
\end{align}
with the correlation between the parameters equal to 0.91. We implement the constraints at 2$\sigma$ level. The formulas for $S$ and $T$ in the IDM can be found for example in ref.~\cite{Swiezewska:2012yuk} (see also references therein). It is important to note that the constraints come mainly from the $T$ parameter, as $S$ is naturally small. In the case of degenerate $A$ and $H^{\pm}$ parameter $T$ vanishes, so the electroweak measurements do not constrain this scenario. 
\item[Relic density of DM.] The current constraints from the Planck experiment give~\cite{Planck}
\be
0.1118 < \Omega_{\mathrm{DM}} h^2 < 0.1280\quad(\mathrm{at}\ 3\sigma). \label{omega}
\ee
This constrains the parameters of the IDM, see refs.~\cite{LopezHonorez:2006,Dolle:2009, Honorez:2010,Sokolowska:2011,Gustafsson:2012,Krawczyk:2013jhep}. Below we will not perform a~scan of the parameter space with the constraint~(\ref{omega}) but we will comment on the consistency of our results and the relic density constraints.
\end{description}

The constraints coming from LEP measurements are important for lighter inert scalars (masses below $\mathcal{O}$(100~GeV)) so we do not consider them here. 

We will start from analysing the case with degenerate $A$ and $H^{\pm}$, as was described in the~previous section. We will examine the regions in the $(M_H,\ M)$ plane where the EWSB minimum is stable/metastable/unstable, and confront them with other constraints. We underline that we do not consider the behaviour of the potential at large field values here, we are only interested in the~stability around the~EW scale. 

\begin{figure}[h]
\center
\includegraphics[width=.45\textwidth]{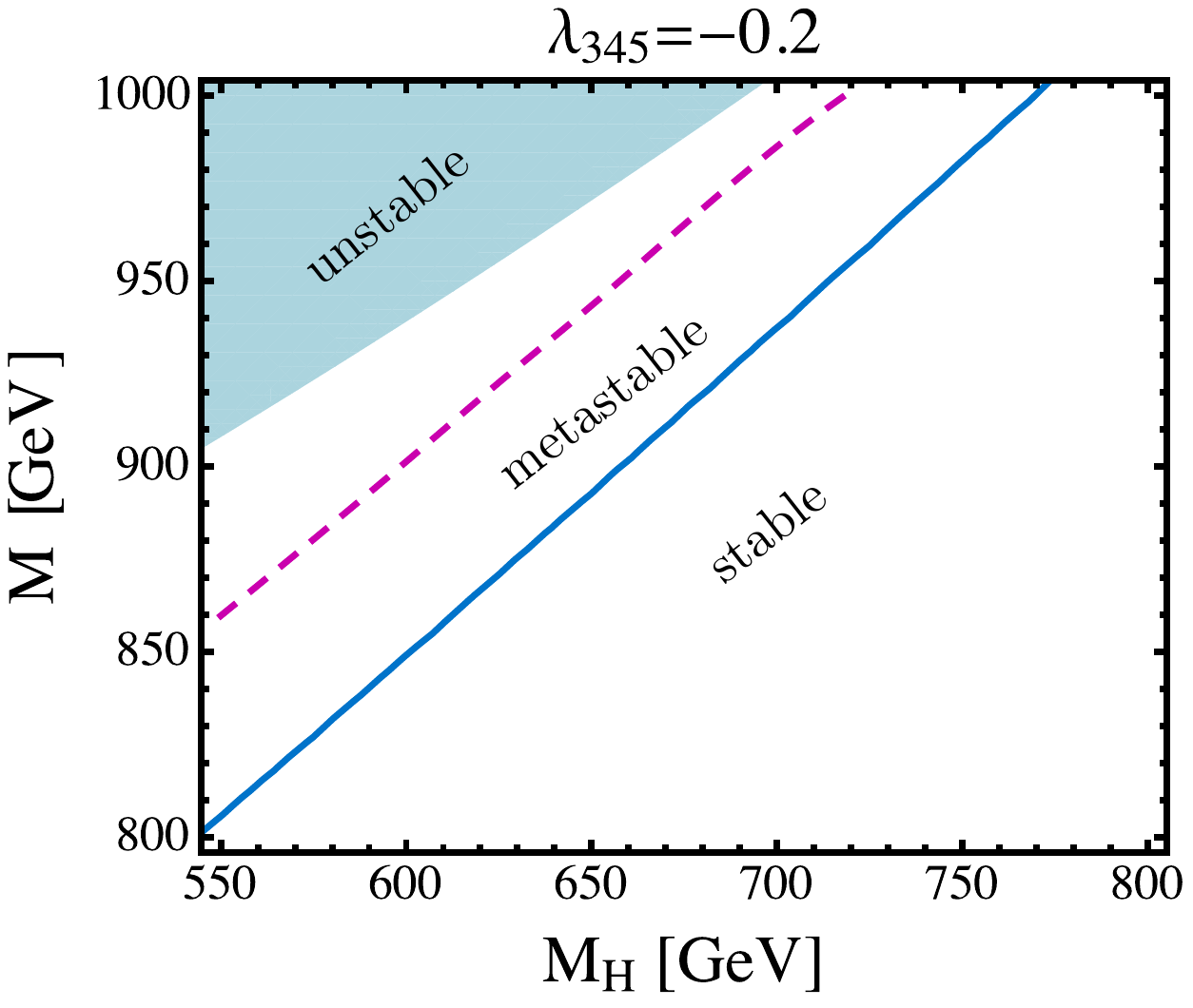}\hspace{.5cm}
\includegraphics[width=.45\textwidth]{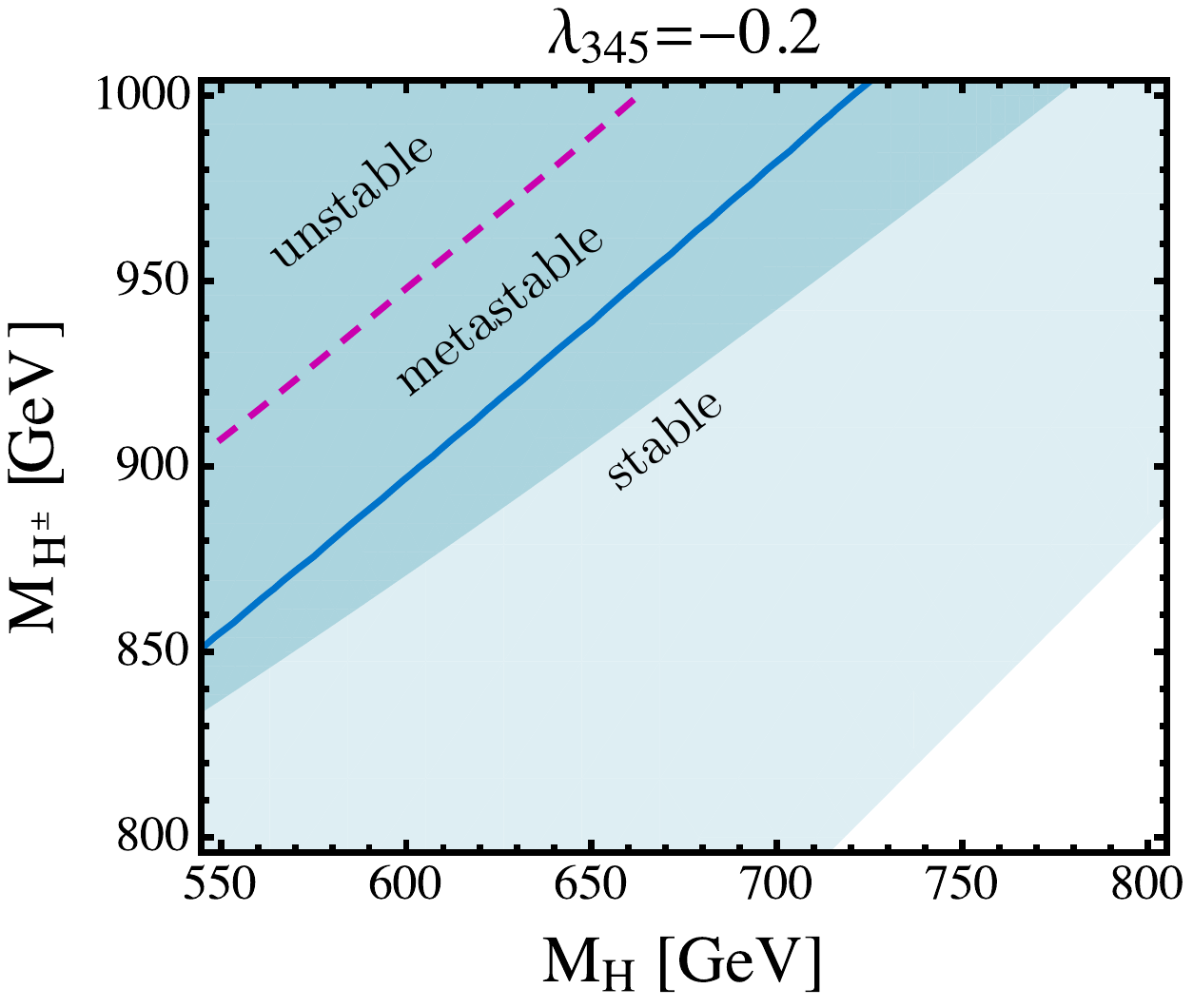}
\caption{Stability/metastability/instability regions for the case with $\lczp=-0.2$, and $M_A=\m=M$ (left panel) or $M_A=M_H+1\g$ (right panel). The solid line denotes the boundary between stable and metastable vacua, the dashed line is the boundary between the metastable and unstable region. The dark shaded region is excluded by unitarity, and the light shaded region is excluded by EWPT. EWPT do not constrain the case with $M_A=\m$.\label{fig:boundaries}}
\end{figure}

The results can be seen in figure~\ref{fig:boundaries} (left panel), the solid line represents the region where $\veff(v)=0$, i.e. the boundary between stability and metastability region. Along the  dashed line $\tau=1$ (in the units of $T_U$) so it is the~boundary between the metastable and unstable vacua. The shaded region is excluded by perturbative unitarity. Since $M_A=\m$ the EWPT  do not introduce new constraints. The parameter $\lczp$ is fixed to $-0.2$. We checked that changing $\lczp$ within the range that is favoured by the relic density constraints ($-0.3\lesssim \lczp \lesssim 0.3$)~\cite{Krawczyk:2013acta} changes the picture only slightly.

It is clear from figure~\ref{fig:boundaries} that meta- and unstable scenarios are in agreement with unitarity constraints\footnote{If rather big values of $\lb$ were considered, the meta- and unstable scenarios could be excluded by unitarity.} and EWPT, as was discussed before. However, for an unstable vacuum to appear, the~splitting between $M_H$ and $M$ has to be large, at the level of 300 GeV. This cannot be reconciled with the relic density constraints --- the heavy DM needs coannihilation with other scalars to develop the correct relic density and the mass splitting among dark scalars must be small~\cite{Hambye:2009}.

Let us then consider a case where $H$ and $A$ are quasi-degenerate (we assume $M_A=M_H+1\g$) to allow for coannihilation processes. Figure~\ref{fig:boundaries} (right panel)  shows the boundaries between regions with vacua of different properties, the coding is the same as in the left panel. Once more we fix $\lczp=-0.2$, and small changes in $\lczp$ do not alter the picture significantly. In this case we have to take into account the EWPT constraints. The light shaded region is excluded by constraints on $S$ and $T$ (it overlaps with the region excluded by unitarity). Unitarity and EWPT exclude the~scenarios where metastability or instability can occur. 

Therefore we conclude that the metastability or instability scenarios within the IDM with heavy scalars cannot be reconciled with theoretical and experimental constraints.

%~~~~~~~~~~~~~~~~~~~~~~
\section{Conclusions\label{sec:conclusions}}
%~~~~~~~~~~~~~~~~~~~~~~

In this work we analysed the impact of new scalar particles on the structure of effective potential of the~IDM around the EW scale.

We showed that the new scalars can have a striking effect on the effective potential. They can turn the maximum of the effective potential at $\f=0$ into a minimum, and moreover change the~energy of the~EWSB minimum in such a way that it becomes only a local one. This gives rise to unstable or metastable EWSB minimum, and the source of instability is around the EW scale.  Our analysis was performed for the IDM but similar effects may be observed in other extensions of the~SM. This shows that it is not enough to consider the behaviour of the effective potential or running coupling constants at large field values. Introduction of new fields can modify the effective potential at low energies and one has to check what effect such modifications have on vacuum stability.

For the particular case of the IDM we checked that the metastability/instability scenario is not a threat since the region where it is realised cannot be reconciled with perturbative unitarity, EWPT and the DM relic abundance measurements by the Planck experiment. 

\paragraph{Note added} At the final stage of preparation of this manuscript  ref.~\cite{Khan:2015} appeared in which vacuum stability in the~IDM is analysed. In~contrast to our work, the focus of this paper is on the~high-scale vacuum stability. In~ref.~\cite{Khan:2015} it was confirmed that additional scalars improve the~running of Higgs self-coupling  and it was shown that even if a new minimum is formed at large energy scales, the lifetime of the vacuum is longer than in the SM.

%~~~~~~~~~~~~~~~~~~~~~
\acknowledgments
%~~~~~~~~~~~~~~~~~~~~~
We are very grateful to M. Krawczyk for suggestions for the analysis, discussions and carefully reading the manuscript, and  V.~Branchina for discussions about metastability issues. We would also like to thank P.~Chankowski and G.~Gil for their help and sharing materials, and P.~M.~Ferreira,  H.~Haber, M.~Lewicki, E.~Messina, P.~Olszewski, and D.~Soko{\l}owska for fruitful discussions. This work was supported by the he Polish National Science Centre grant PRELUDIUM under the~decision number DEC-2013/11/N/ST2/04214.

%~~~~~~~~~~~~~~~~~~~~
\appendix
%~~~~~~~~~~~~~~~~~~~~

\section{Self-energy and tadpole of the Higgs boson in the IDM\label{app}}

The Higgs boson self-energy and the tadpole were computed using dimensional regularisation and can be expressed in terms of basic Passarino-Veltman integrals~\cite{Passarino:1978}, defined as follows
\begin{align}
a(m) &= \int \frac{d^D k}{(2\pi)^D} \frac{i\mu^{\epsilon}}{k^2 - m^2 + i \epsilon},\nonumber\\*
b_0(p^2, m_1, m_2) &= \int \frac{d^D k}{(2\pi)^D} \frac{i\mu^{\epsilon}}{(k^2 - m_1^2 + i \epsilon)\left[(p-k)^2-m_2^2+i\epsilon\right]}.\nonumber
\end{align}
Using the standard Feynman parametrisation, and expansion in $\epsilon$, the functions can be evaluated, up to terms vanishing for $\epsilon \to 0$ as
\begin{align}
a(m) &= -\frac{m^2}{(4\pi)^2}\left(\frac{2}{\epsilon}  -\gamma_E+ \log (4\pi \mu^2)-\log {m^2} +1 \right),\nonumber\\
b_0(p^2,m_1, m_2) &= -\frac{1}{(4\pi)^2}\left(\frac{2}{\epsilon}  -\gamma_E + \log (4\pi \mu^2) - \int_0^1 dx \log \Delta\right),\nonumber
\end{align}
where $\Delta = -x(1-x) p^2 + x m_1^2 +(1-x) m_2^2$.

We also introduce a non-standard $a^b$ and $b_0^b$ functions, which will be useful for the bosonic loops
\begin{align}
3 a^b(m) &=(D-1)a(m)=-\frac{3m^2}{(4\pi)^2}\left(\frac{2}{\epsilon}  -\gamma_E+ \log (4\pi \mu^2)-\log {m^2} +\frac{1}{3} \right),\nonumber\\
4b_0^b(p^2,m_1, m_2) &=Db_0(p^2,m_1, m_2)= -\frac{4}{(4\pi)^2}\left(\frac{2}{\epsilon}  -\gamma_E + \log (4\pi \mu^2) - \int_0^1 dx \log \Delta - \frac{1}{2}\right).\nonumber
\end{align}
They differ from the original ones only by the finite part.\footnote{These functions would not appear if we used dimensional reduction (DRED) instead of dimensional regularisation (DREG).}

The Higgs tadpole in the IDM is given by
\begin{align}
-i\mathcal{T}=&-i\left[\frac{3}{2} \lambda_1 a(M_h)+ \frac{3}{2} g^2 a^b(M_W) + \frac{3}{4} (g^2 + g'^2)a^b(M_Z)
  - 6y_t^2 a(M_t) - 6 y_b^2 a (M_b)\right.\nonumber\\
&  +\left.\lc a(\m)+\frac{1}{2}\lczp a(M_H) +\frac{1}{2} \lczp^- a(M_A)\right]v.\nonumber
\end{align}

The Higgs self-energy is given by
\begin{align}
\Sigma(p^2)&=\frac{g^2}{4M_W^2}\left[16M_W^4b_0^b(p^2, M_W, M_W)+\left(p^4-4p^2M_W^2-4M_W^4\right)b_0(p^2, M_W, M_W)\right]\nonumber\\
&+ \frac{g^2+g'^2}{8M_Z^2}\left[16M_Z^4b_0^b(p^2, M_Z, M_Z)+\left(p^4-4p^2M_Z^2-4M_Z^4\right)b_0(p^2, M_Z, M_Z)\right]\nonumber\\
&+ b_0(p^2,0,0)\left(-\frac{g^2}{4M_W^2}p^4 - \frac{g^2+g'^2}{8M_Z^2}p^4 + \frac{3}{8}\frac{ g^2M_h^4}{M_W^2}\right)\nonumber\\
&-\frac{g^2p^2}{2M_W^2}a(M_W) +\frac{3g^2}{2} a^b(M_W)-\frac{(g^2+g'^2)p^2}{4M_Z^2}a(M_Z) +\frac{3(g^2+g'^2)}{4} a^b(M_Z)\nonumber\\
&+\frac{9}{8}g^2 \frac{M_h^4}{M_W^2}b_0(p^2,M_h,M_h)+\frac{3}{8} g^2 \frac{M_h^2}{M_W^2} a(M_h)\nonumber\\
&-\frac{3g^2}{2M_W^2}M_t^2\left[2a(M_t)+(-p^2+4M_t^2)b_0(p^2,M_t,M_t)\right]\nonumber\\
&-\frac{3g^2}{2M_W^2}M_b^2\left[2a(M_b)+(-p^2+4M_b^2)b_0(p^2,M_b,M_b)\right]\nonumber\\
&+\lc a(\m)+\frac{1}{2}\lczp a(M_H) +\frac{1}{2} \lczp^- a(M_A)\nonumber\\
&+(\lc v)^2 b_0\left(p^2,\m,\m\right) + \frac{1}{2} (\lczp v)^2 b_0\left(p^2,M_H,M_H\right) + \frac{1}{2}(\lczp^- v)^2b_0\left(p^2,M_A,M_A\right).\nonumber
\end{align}

\bibliography{Swiezewska}
\end{document}